\documentclass[12pt, onecolumn, draftclsnofoot]{IEEEtran}
\usepackage{pdfsync}
\usepackage{algorithm, algorithmic}
\usepackage{epsfig, graphics, color}
\usepackage{amsmath, amssymb, amsfonts, amsbsy, mathrsfs, bm}
\usepackage{arydshln, multirow}
\usepackage{cases, enumerate, fancyvrb}
\usepackage{theorem}
\usepackage[compress]{cite}
\usepackage{psfrag}
\usepackage[normalem]{ulem}
\usepackage[margin=0.9in]{geometry}
\DeclareMathAlphabet{\mathsfsl}{OT1}{cmss}{m}{sl}

\ifCLASSOPTIONcompsoc
 \usepackage[tight, normalsize, sf, SF]{subfigure}
\else
 \usepackage[tight, footnotesize]{subfigure}
\fi

\newtheorem{theorem}{\bf Theorem}
\newtheorem{proposition}{\bf Proposition}
\newtheorem{lemma}{\bf Lemma}

\newtheorem{assumption}{\bf Assumption}

\newcounter{rcounter}
\newenvironment{remarks}{
	\begin{list}{\textit{Remark} \arabic{rcounter}:~}{
    \setcounter{enumi}{\value{rcounter}}
    \usecounter{rcounter}
    \setcounter{rcounter}{\value{enumi}}
    \setlength\labelwidth{0in}
    \setlength\labelsep{0in}
    \setlength\leftmargin{0in}
    \setlength\listparindent{0in}
    \setlength\itemindent{15pt}}
}{
	\end{list}
}

\newcommand{\bp}{\mathbf{p}}
\newcommand{\br}{\mathbf{r}}
\newcommand{\bs}{\mathbf{s}}
\newcommand{\bu}{\mathbf{u}}
\newcommand{\bv}{\mathbf{v}}
\newcommand{\bw}{\mathbf{w}}
\newcommand{\bx}{\mathbf{x}}
\newcommand{\by}{\mathbf{y}}

\newcommand{\bo}{\mathbf{0}}

\newcommand{\bbeta}{\bm{\beta}}

\newcommand{\bmeta}{\bm{\eta}}

\newcommand{\bphi}{\bm{\phi}}
\newcommand{\brho}{\bm{\rho}}
\newcommand{\bPhi}{\bm{\Phi}}
\newcommand{\Rho}{\mathrm{P}}
\newcommand{\bRho}{\bm{\Rho}}
\newcommand{\bSigma}{\bm{\Sigma}}
\newcommand{\bGamma}{\bm{\Gamma}}
\newcommand{\bXi}{\bm{\Xi}}
\newcommand{\bPi}{\bm{\Pi}}

\newcommand{\bbA}{\bar{\mathbf{A}}}
\newcommand{\bbB}{\bar{\mathbf{B}}}
\newcommand{\bbC}{\bar{\mathbf{C}}}
\newcommand{\bbD}{\bar{\mathbf{D}}}

\newcommand{\bA}{\mathbf{A}}
\newcommand{\bB}{\mathbf{B}}
\newcommand{\bC}{\mathbf{C}}
\newcommand{\bD}{\mathbf{D}}
\newcommand{\bE}{\mathbf{E}}
\newcommand{\bF}{\mathbf{F}}

\newcommand{\bH}{\mathbf{H}}
\newcommand{\bI}{\mathbf{I}}
\newcommand{\bJ}{\mathbf{J}}
\newcommand{\bK}{\mathbf{K}}
\newcommand{\bL}{\mathbf{L}}
\newcommand{\bQ}{\mathbf{Q}}
\newcommand{\bU}{\mathbf{U}}
\newcommand{\bV}{\mathbf{V}}

\newcommand{\dd}{\mathrm{d}}
\newcommand{\re}{\mathrm{e}}

\newcommand{\T}{\mathcal{T}}
\newcommand{\D}{\mathcal{D}}

\newcommand{\stB}{\mathcal{B}}
\newcommand{\stS}{\mathcal{S}}

\newcommand{\Pb}{\mathsf{P}_b}
\newcommand{\SNRb}{\mathsf{SNR}_b}
\newcommand{\Pn}{\mathsf{P}_0}
\newcommand{\Wb}{\mathsf{W}_b}
\newcommand{\Tb}{\mathsf{T}_b}

\newcommand{\R}{\mathbb{R}}

\newcommand{\ri}{\mathrm{i}2\pi}

\newcommand{\EE}[2]{{\mathbb{E}_{#1}\left\{ #2 \right\}}}

\newcommand{\abs}[1]{{\left\lvert #1 \right\rvert}}
\newcommand{\norm}[1]{{\left\lVert #1 \right\rVert}}

\newcommand{\rec}[1]{{\mathrm{rect}\left( #1 \right)}}
\newcommand{\der}[2]{\frac{\partial #1}{\partial #2}}

\newcommand{\derdd}[3]{\frac{\partial^2 #1}{\partial #2 \partial #3}}
\newcommand{\ofrac}[1]{{\frac{1}{#1}}}
\newcommand{\condp}[2]{p \left(\left. #1 \right| #2 \right)}

\newcommand{\Real}[1]{\mathcal{R}\mathrm{e}\left\{#1\right\}}

\newcommand{\blue}[1]{{{\color{blue} #1}}}

\title{Modified CRB for Location and Velocity Estimation using Signals of Opportunity}
\author{Mei Leng, Wee Peng Tay, Chong Meng Samson See, Sirajudeen Gulam Razul, and Moe Z. Win
\thanks{M. Leng and W.P. Tay are with the Nanyang Technological University, Singapore. (e-mail:$\{$lengmei,wptay$\}$@ntu.edu.sg).}
\thanks{C.M. Samson See and S.G. Razul are with TL@NTU, Singapore. (e-mail: samsonsee,ESirajudeen@ntu.edu.sg).}
\thanks{Moe Z. Win is with the Laboratory for Information and Decision Systems (LIDS), Massachusetts Institute of Technology (e-mail: moewin@mit.edu).}
\thanks{This research is supported by the DRTech-TL@NTU grant TL-9010100261-04.}}

\begin{document}
\maketitle

\begin{abstract}
We consider the problem of localizing two agents using signals of opportunity from beacons with known positions. Beacons and agents have asynchronous local clocks or oscillators with unknown clock skews and offsets. We model clock skews as random, and analyze the biases introduced by clock asynchronism in the received signals. By deriving the equivalent Fisher information matrix for the modified Bayesian Cram\'er-Rao lower bound (CRLB) of agent position and velocity estimation, we quantify the errors caused by clock asynchronism. We propose an algorithm based on differential time-difference-of-arrival and frequency-difference-of-arrival that mitigates the effects of clock asynchronism to estimate the agent positions and velocities. Simulation results suggest that our proposed algorithm is robust and approaches the CRLB when clock skews have small standard deviations.
\end{abstract}


\section{Introduction} \label{sect:intro}
The availability of reliable, real-time and high-accuracy location-awareness is essential for current and future wireless applications \cite{Win2011}. For example, the positions of mobile terminals are indispensable for location based services, search and rescue operations cannot be fulfilled without accurate navigation, and a large set of emerging wireless sensor network applications \cite{Tay2008,Kreidl2010,Tay2009} requires sensor locations to meaningfully interpret the collected data. Various wireless devices have been used in these applications, and we will use  ``\textit{device}'' and ``\textit{agent}'' interchangeably in this paper. A common localization method is the use of Global Positioning System (GPS). However, GPS signals are generally limited to areas with a clear sky view, and do not penetrate well through obstacles, which makes it impractical for use in indoor and urban environments. As a promising alternative to GPS localization, the exploitation of signals of opportunity (SOOP) has been attracting much interest recently \cite{Tian2008, Moghtadaiee2010, Robinson2012}. SOOP are public signals transmitted for various non-localization applications, including AM/FM radio signals \cite{Fang2009,Moghtadaiee2011}, digital television signals \cite{Serant2010}, and cellular communication signals \cite{Bshara2011}. These signals conform to well-established standards and can be easily detected in most urban areas. We call their transmitters ``\textit{beacons}'' and their locations can be obtained a priori. 

Various measurements can be taken from SOOP for navigation purposes, including received-signal-strength (RSS), angle-of-arrival (AOA), time-difference-of-arrival (TDOA) and frequency-difference-of-arrival (FDOA). Positioning based on RSS level is relatively cost-effective solution since the signal power function is available in most mobile devices and hence localization can be achieved without any hardware modification. However, it generally requires assuming that the transmitted power is known and its accuracy is insufficient due to the complex propagation model for signal power attenuation. The localization based on AOA on the other hand requires large antenna arrays for angle measurement, which prohibits its usage on smart mobile devices, and its accuracy deteriorates rapidly when agents move away from the beacon. Compared with the first two methods, TDOA and FDOA provide an attractive alternative, since they can removes the ambiguity caused by unknown transmit time or unknown signal waveforms, and this is especially important for using SOOP where agents usually have no access to the transmit time and transmitted waveform. Therefore, we focus on TDOA and FDOA measurements in this paper.

Since SOOP are not designed for geo-localization, various challenges are encountered when utilizing such signals for positioning. The first challenge stems from the fact that most beacons are passive transmitter and will not cooperate with agents. Therefore, each agent has no knowledge on important information such as transmit power, transmit time, and signal waveforms. Traditionally, an extra reference agent is employed, it is assumed to have a known position and cooperates with the agent-to-be-localized to extract range measurements from SOOP \cite{Martin2009}. However, it requires that both agents receive the same transmission and the maintenance of such a reference device can be costly, especially when the agent is moving and the number of agents increases. In this paper, we consider the localization of two moving agents whose locations are both unknown, and it serves as a building block for the problem of network localization \cite{Win2011}.

The most critical challenge for localization using SOOP is synchronization \cite{Moghtadaiee2010}. In order to obtain reliable timing information, it is essential that signals from all beacons are synchronized and each received SOOP is also processed in a synchronized manner. Existing methods usually assume that the clocks of beacons and agents are synchronized (see \cite{Bshara2011, Yeredor2011} and references therein). However, clock synchronization is difficult to achieve and maintain in practice \cite{Leng2011, Leng2011a}, and even worse situations are encountered in practice where some beacons, such as GSM base stations, are not synchronized. Unlike GPS signals that are generated by satellite atomic oscillators with extremely small skews (less than $10^{-11}$, \cite{Cheng2005}), SOOP are usually generated by beacons with less perfect oscillators that have clock skews varying from $10^{-8}$ to $10^{-4}$ \cite{Cristian1989, Agilent1997}, which results in an accumulated clock offset up to 0.1 ms for one second. Therefore, by deriving a fundamental limit for localization in presence of clock asynchronism in this paper, we are interested to investigate how much asynchronous clocks will affect localization accuracy and it provides insight for designing localization algorithms in practice.

Source localization algorithms using a set of asynchronous static agents were proposed in \cite{Li2004} and \cite{Li2006}. The source is assumed to emit a sequence of short pulses, and each pulse has a period that can be measured by a known number of clock ticks. TDOA measurements are made by counting the number of received pulses, which requires specially designed pulse waveforms. A more general case was investigated in \cite{Yan2008} where all beacons and agents have asynchronous clocks. They assume that each agent can estimate the time-of-arrival (TOA) with respect to a beacon. The TDOA between two agents is then obtained by subtracting one TOA from another. In practice, the direct estimation of TOA is however difficult, since agents neither know the transmitted waveform nor are able to decode the received signal for timing information. In order to apply real-time localization, two agents must exchange their received signal and obtain local TDOA measurement by correlation \cite{Moghtadaiee2010}. The distortions from communication and local processing are therefore inevitable and must be taken into consideration. 

In this paper, we investigate the localization and velocity estimation of moving agents using arbitrary narrowband SOOP. We assume that agents know the nominal carrier frequencies of the beacons but have no knowledge of their waveforms. We also assume that each agent or beacon has a local oscillator that controls its clock, and we use the terms clock and oscillator interchangeably. We are interested to quantify the fundamental performance limit of agent location and velocity estimation in the presence of clock asychronism, and to develop estimation algorithms that mitigate its effects. Our main contributions are the following.
\begin{itemize}
	\item We analyze the biases introduced by asynchronous clocks in the beacons and agents, and derive expressions for the received signals at each agent (cf.\ Theorem \ref{theorem:signals}). We show that the time and frequency offsets in the received signals are corrupted by beacon and agent clock offsets and skews. Our analysis includes the scenario studied in \cite{Yan2008} as a special case.
	
	\item We derive the fundamental limits for agent location and velocity estimation using SOOP from asynchronous beacons (cf.\ Theorem \ref{theorem:crlb}). The CRLB for localization and tracking have been extensively studied in the literature, but most works are based on the statistical characteristics of signal metrics like TDOA and FDOA \cite{So2008}. As pointed out in \cite{Shen2010}, signal metrics depend heavily on the specific measurement method, which may discard useful information during processing. We treat the agent clock skews as nuisance random variables, and we derive an approximate equivalent Fisher information matrix (EFIM) associated with the modified Bayesian CRLB for agent locations and velocities, directly from the received signals. We show that the approximate EFIM does not depend on agent and beacon clock offsets, which suggests that there exists estimation procedures that does not require a priori knowledge of these quantities.
\end{itemize}

The rest of this paper is organized as follows. In Section \ref{sect:system}, we present the system model. In Section \ref{sect:crlb}, we derive the Fisher information matrix for the modified Bayesian CRLB based on the received signals at each agent. For the reader's ease of reference, we summarize some commonly used notations in Table \ref{table:notation}.

\begin{table}[!t]
    \caption{Commonly Used Notations}\label{table:notation}
    \centering
		\vspace{-.3cm}
    \begin{tabular}{|c||l|}
    \hline
    Notations & Definition \\
    \hline
    $\stS$, $\stB$ & set of agents and beacons, respectively \\
    \hline
    $\beta_m$, $\Omega_m$ & clock skew and clock offset of $m\in\stS\cup\stB$, respectively \\
    \hline
    $\T_{j,b}$, $\D_{j,b}$ & propagation delay and nominal Doppler shift from beacon $b$ to agent $j$, respectively \\
    \hline
    $\bu_{j,b}$, $\bw_{j,b}$ & direction vectors between agent $j$ and beacon $b$ defined in \eqref{D} and \eqref{w}, respectively \\
    \hline
    $\hat{\tau}_b$, $\hat{\xi}_b$ & TDOA and FDOA estimates respectively at $S_1$ using signals of beacon $b$ \\
    \hline
    $\bF_{\bx}$ & Fisher information matrix (FIM) of the modified Bayesian CRLB for $\bx$ \\
    \hline
		$\bF_{\re,\bs}$ & equivalent Fisher information matrix (EFIM) for agent location and velocity estimates \\
    \hline
    $\Pb$, $\Wb$, $\Tb$ & energy, RMS bandwidth, and RMS integration time of the signal transmitted from beacon $b$, respectively \\
    \hline
    $\lambda_b$, $\epsilon_b$ & contants related to beacon $b$'s signal characteristics and defined in \eqref{chi-omega}\\
    \hline
    \end{tabular}
		\vspace{-.8cm}
\end{table}

\section{System Model}\label{sect:system}
\begin{figure}[!t]
  \centering
  \includegraphics[width=0.35\textwidth]{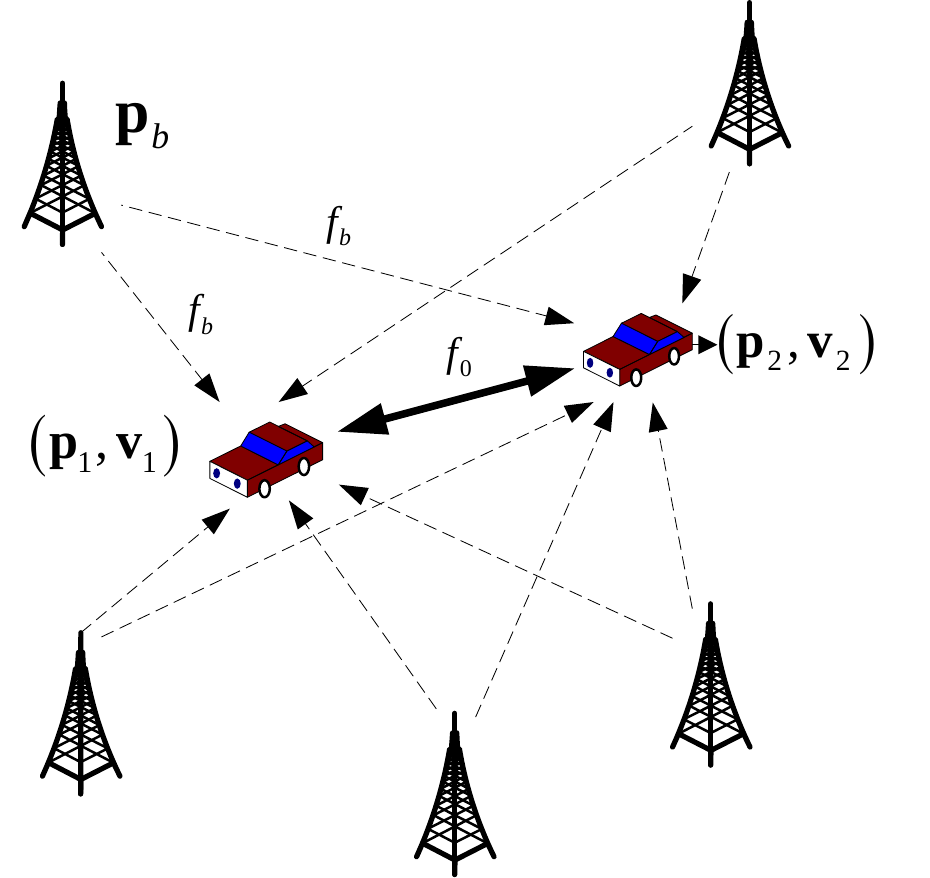}
	\vspace{-.5cm}
  \caption{Using signals from beacons to estimate agent locations and velocities.}
  \label{Figure-Environment}
	\vspace{-1.2cm}
\end{figure}

We consider the problem of localizing two agents using signals from a set $\stB$ of $N$ beacons. An example scenario is shown in Figure \ref{Figure-Environment}. The beacon $b \in \stB$ has a known fixed position $\bp_b$, and it broadcasts a narrowband signal at a nominal carrier frequency $f_b$. Two agents, $S_1$ and $S_2$ are at unknown locations $\bp_1$ and $\bp_2$, and moving with unknown velocities $\bv_1$ and $\bv_2$ respectively. Denote the set of agents as $\stS=\{1,2\}$. Our goal is to estimate the locations $\bp_j$ and velocities $\bv_j$ for $j\in\stS$, using SOOP from the beacons.

We suppose that each beacon $b$ has a local oscillator operating with clock skew $\beta_b$ and clock offset $\Omega_b$, so that its local time $t_b(t)$ with respect to (w.r.t.) a universal standard time $t$ is given by \cite{Sundararaman2005}
\begin{align}
t_b(t) = \beta_bt + \Omega_b.
\end{align}
The clock skew $\beta_b$ characterizes the clock drift rate and the clock offset $\Omega_b$ characterizes any clock errors and clock drifts accumulated up to time $t$. We approximate the clock offset $\Omega_b$ to be constant over our observation period, and its value generally ranges from $-10$ms to $10$ms \cite{Cristian1989}. Similarly, each agent $j$ has a local oscillator with local time given by $t_j(t) = \beta_jt + \Omega_j$. 

The rate at which an oscillator drifts depends on various random quantities like its quality, power level, temperature and other environmental variables. Since clock skew is non-negative, we assume that agent and beacon clock skews are independent Gamma random variables with mean 1 and standard deviation $\sigma_{\beta_m}$, for $m \in\stS\cup\stB$. We also assume that these random variables do not depend on the beacon or agent positions and velocities. The expected value of clock skews is assumed to be $1$ because clocks generally drift slowly w.r.t. the standard time and its standard deviation $\sigma_{\beta_m}$ varies between $10^{-6}$ and $10^{-3}$ \cite{Cristian1989, Lewis1991}. It will be obvious later that our analysis is easily generalizable to other distributions \cite{Harris2001} as long as the quantities $\mathbb{E}\{1/\beta_m\}$ and $\mathbb{E}\{1/\beta_m^2\}$ exist and are finite.

We assume that each agent has no prior knowledge of the signal waveform transmitted by beacons. However, agents know the \textit{nominal} carrier frequencies used by the beacons and the positions of the beacons. We also assume that signals from different beacons can be distinguished at agents. Sensors forward its received signals to each other by communicating over a wireless channel with nominal passband frequency $f_0$. To perform self-localization, each agent obtains TDOA and FDOA measurements by cross-correlating signals received by itself and by the other agent. Since local signal processing, including signal generation and sampling, depends on the local oscillator, the TDOA and FDOA measurements obtained at each agent will be distorted by their asynchronous oscillators. In the following, we first analyze how such distortions affect the transmitted and received signals, and then derive closed-form expressions for TDOA and FDOA based on the distorted signals.

\subsection{Delays and Doppler Shifts}\label{sect:channel}
In this section, we briefly describe the wireless channel for communications between beacons and agents. We assume that every wireless channel has a flat fading time-varying impulse response. For each $b\in \stB$, and each agent $S_j$, $j\in\stS$, we suppose that the channel between $b$ and $S_j$ have an impulse response $h_{j,b}(t) = \alpha_{j,b} \delta(t - \tau_{j,b}(t))$ \cite{Gallager2008}. Let the propagation delay between $b$ and $S_j$ be
\begin{align}
\T_{j,b} & = \frac{\norm{\bp_j - \bp_b}}{c}, \label{T}
\end{align}
where $c$ is the speed of light, and let the nominal Doppler shift observed at $S_j$ be
\begin{align}
\D_{j,b} & = -\frac{f_b}{c}\bv_j^T\underbrace{\frac{\bp_j - \bp_b}{\norm{\bp_j - \bp_b}}}_{\triangleq \bu_{j,b}}, \label{D}
\end{align}
where $\bu_{j,b}$ is the normalized direction from $S_j$ to beacon $b$. We approximate the propagation time from $b$ to $S_j$ as $\tau_{j,b} (t) \approx \T_{j,b} - \D_{j,b} t/f_b$. Similarly, we approximate the propagation time from agent $S_2$ to $S_1$ as $\tau_{1,2}(t) \approx \T_{1,2} - \D_{1,2}t/f_0$, where $\T_{1,2} = \norm{\bp_1 - \bp_2}/c$ and $\D_{1,2} = -f_0(\bv_1-\bv_2)^T\bu_{1,2}/c$ is the Doppler shift observed at $S_1$.

\subsection{Received Signals at Sensors}
\begin{figure}[!t]
  \centering
  \includegraphics[width=0.5\textwidth]{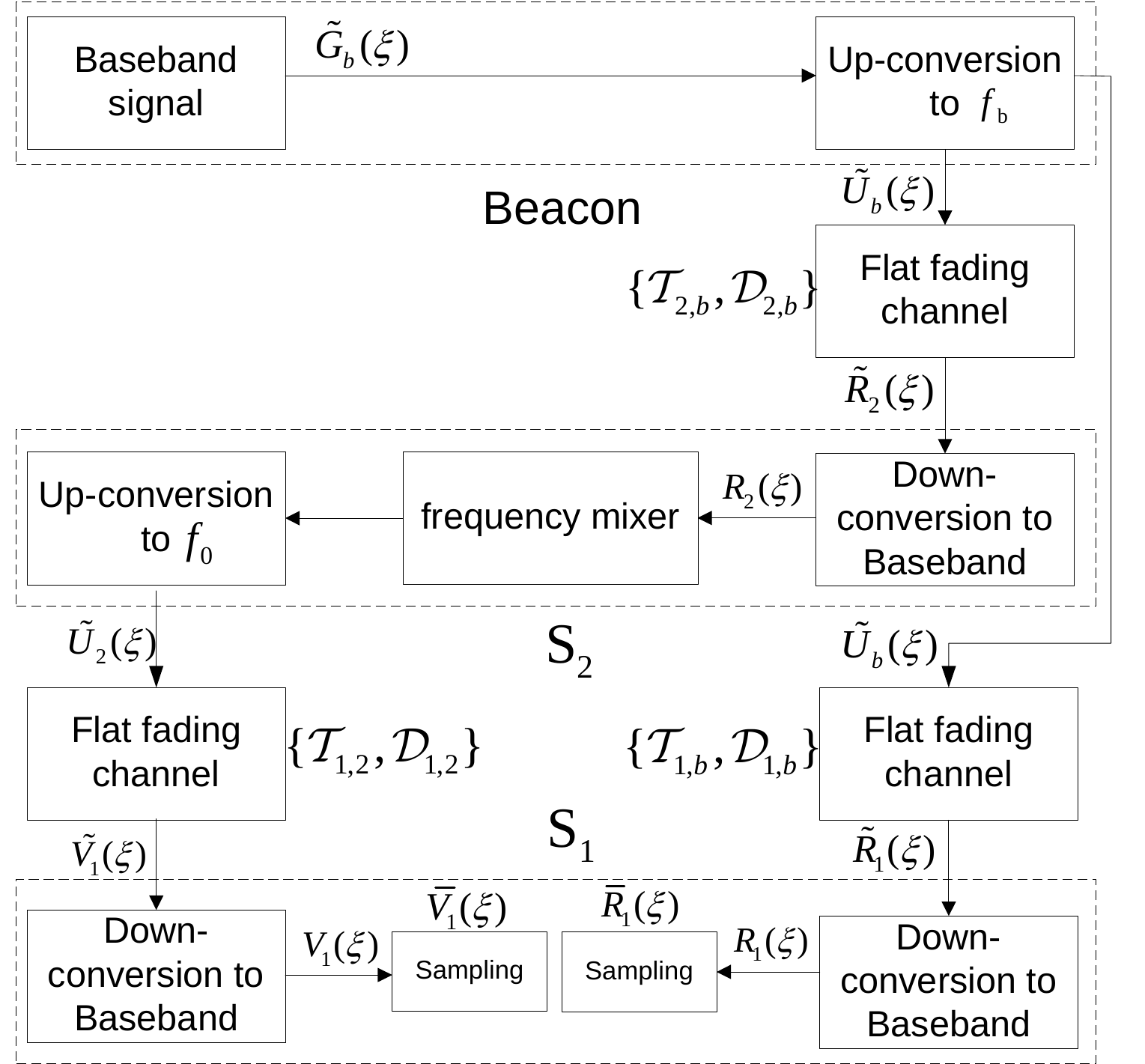}
	\vspace{-.5cm}
  \caption{Block diagram for communications between beacons and agents.}
  \label{Figure-System}
	\vspace{-.5cm}
\end{figure}
Suppose that beacon $b$ generates a nominal baseband signal $g_b(t)$. Because of clock skew and offset at the beacon, the actual baseband signal is $\tilde{g}_b(t) = g_b(\beta_b t+\Omega_b)$, which is then up-converted to the nominal passband frequency $f_b$ at the beacon for transmission, and the actual passband frequency may differ from $f_b$ due to the clock skew of beacon $b$. Since we do not assume that agents know the nominal baseband signal $g_b(t)$ used by the beacon, it suffices to consider only the signal $\tilde{g}_b(t)$. However, to make the connections with TOA methods and scenarios where $g_b(t)$ is a known pilot signal clear, we have chosen to explicitly characterize signals in terms of $g_b(t)$ (cf.\ Remarks \ref{remark:syncB} and \ref{remark:syncS} later). Let $G_b(\xi)$ be the Fourier transform of $g_b(t)$.

The agent $S_1$ receives the signal from beacon $b$, and then performs demodulation and sampling (see Figure \ref{Figure-System}). Each step involves signal processing using the local oscillator, which introduces errors due to its clock skew and offset. We derive the signal at each processing step {in Proposition \ref{proposition:signals}, whose proof is given in Appendix \ref{appdx-signal}. Let $\mathrm{i} = \sqrt{-1}$, $\gamma_{1,2}= 1+\D_{1,2}/f_0$, and $\gamma_{j,b} = 1+\D_{j,b}/f_b$.

\begin{proposition}\label{proposition:signals}
Suppose that beacon $b\in \stB$ transmits a signal with baseband representation $\tilde{G}_b(\xi)$ to agent $S_j$ over a flat fading channel. We have the following.
\begin{enumerate}[(i)]
\item\label{it:gen} At beacon $b$, the baseband signal $\tilde{g}_b(t)$ has positive frequency part
$$\tilde{G}_b(\xi) = \frac{1}{\beta_b}G_b(\xi/\beta_b)\exp\{\ri\xi\Omega_b/\beta_b\}.$$

\item\label{it:up} At beacon $b$, the transmitted passband signal $\tilde{u}_b(t)$ has positive frequency part
$$\tilde{U}_b(\xi) = \tilde{G}_b(\xi - f_b\beta_b)\exp\{\ri f_b\Omega_b\}.$$

\item\label{it:prop} At $S_j$, the passband signal received from beacon $b$, $\tilde{r}_{j,b}(t)$, has positive frequency part
$$\tilde{R}_{j,b}(\xi) = \ofrac{\gamma_{j,b}}\tilde{U}_b(\xi/\gamma_{j,b})\exp\{-\ri \xi \T_{j,b}/\gamma_{j,b}\}.$$

\item\label{it:down} At $S_j$, the signal after down conversion to baseband, $r_{j,b}(t)$, has positive frequency part
$$R_{j,b}(\xi) = \tilde{R}_{j,b}(\xi + f_b\beta_j)\exp\{-\ri f_b\Omega_j\}.$$

\item\label{it:sample} Suppose $S_j$ samples the received signal at Nyquist rate of $1/T$ so that the sampling impulse train is $\sum_{n = -\infty}^{+\infty} \delta(\beta_jt - nT)$. The sampled signal has positive frequency part $R_{j,b}(\beta_j\xi)$.
\end{enumerate}
\end{proposition}

Sensor $S_1$ utilizes both its own received signal and that received by agent $S_2$ to obtain TDOA and FDOA measurements. In practice, agent $S_2$ forwards its received signal to $S_1$ by first doing a frequency translation to passband $f_0$ and then transmitting through the wireless channel $h_{1,2}(t) = \alpha_{1,2} \delta(\gamma_{1,2}t - \T_{1,2})$ (cf.\ Figure \ref{Figure-System}). The received signals at agent $S_1$ after sampling is given in Theorem \ref{theorem:signals}, whose proof is provided in Appendix \ref{appdx-rjvj}.

\begin{theorem} \label{theorem:signals}
Let $g_b(t)$ be the nominal baseband signal to be generated by beacon $b$, and $G_b(\xi)$ be its Fourier transform. The received signal at agent $S_1$ from beacon $b$ has baseband representation
\begin{align}
R_{1,b}(\xi) & \propto G_b\left[(\beta_1\xi - \Upsilon_{1,b})/(\beta_b\gamma_{1,b})\right] \re^{-\ri \xi \frac{\beta_1\Delta_{1,b}}{\beta_b\gamma_{1,b}}}, \label{r1}
\end{align}
where $\Delta_{1,b} = \T_{1,b}\beta_b - \Omega_1\beta_b/\beta_1 - \Omega_b$ and $\Upsilon_{1,b} = f_b(\gamma_{1,b}\beta_b-\beta_1)$. The received signal at agent $S_1$ from agent $S_2$ has baseband representation
\begin{align}
\hspace{-.2cm} V_{1,b}(\xi) & \propto G_b\left[(\beta_1\xi - \Psi_{1,b})/(\beta_b\gamma_{1,2}\gamma_{2,b})\right] \re^{-\ri \xi \frac{\beta_1\Lambda_{1,b}}{\beta_b\gamma_{1,2}\gamma_{2,b}}}, \label{v1}
\end{align}
where $\Lambda_{1,b} = \T_{2,b}\beta_b+\T_{1,2}\gamma_{2,b}\beta_b-\Omega_2\beta_b/\beta_2-\Omega_b$, and $\Psi_{1,b}=f_b\gamma_{1,2}(\gamma_{2,b}\beta_b-\beta_2) + f_0(\gamma_{1,2}\beta_2 - \beta_1)$.
\end{theorem}

\begin{remarks}
\item\label{remark:syncB} Suppose that agents are static. When all beacon clocks are synchronized (i.e., $\beta_b=1$ and $\Omega_b=0$ for $b\in\stB$), the received signal \eqref{r1} reduces to that in \cite{Shen2010}. Assuming further that the signal $g_b(t)$ can be generated locally by agent $S_1$, the TOA estimate obtained by cross-correlating $r_{1,b}(t)$ and $g_b(t)$ is given by $\T_{1,b}-\Omega_1$ as in \cite{Shen2010}.
\item\label{remark:syncS} Suppose that agents are static. The local TOA value at $S_j$ in \cite{Yan2008} is based on time-stamp values and is given by $\T_{j,b}\beta_b-\Omega_1\beta_b/\beta_1-\Omega_b$, which equals to $\Delta_{1,b}$ in Theorem \ref{theorem:signals}. The TDOA between two agents $S_1$ and $S_2$ in \cite{Yan2008} is then given by direct subtraction as $\Delta_{2,b} - \Delta_{1,b}$.
\end{remarks}

\section{Modified CRLB for location and velocity estimation}\label{sect:crlb}
In this section, we derive the fundamental limits for location and velocity estimation using the received signals at $S_1$. Let $T$ be the sampling interval, and $T_{ob}$ be the total observation time. For each beacon $b\in\stB$, let $r_{b}[1:T_{ob}/T]$ and $v_b[1:T_{ob}/T]$ be the sampled sequence of the received signal from $b$ received at agent $S_1$, and the signal from $b$ retransmitted from $S_2$ to $S_1$, respectively. Taking the inverse Fourier transform of \eqref{r1} and \eqref{v1}, we have for each $l = 1, \cdots, T_{ob}/T$,
\begin{align}
r_b[l] & = \underbrace{g_b(\beta_b \gamma_{1,b}t/\beta_1 - \Delta_{1,b})\exp\{-\ri\Upsilon_{1,b}t/\beta_1\}|_{t = lT}}_{\triangleq \mu_b[l]} + \varpi_b^r(t)|_{t = lT}, \label{rjn}\\
v_b[l] & = \underbrace{g_b(\beta_b \gamma_{1,2}\gamma_{2,b}t/\beta_1 - \Lambda_{1,b})\exp\{-\ri\Psi_{1,b}t/\beta_1\}|_{t = lT}}_{\triangleq \theta_b[l]} + \varpi_b^v(t)|_{t = lT}. \label{vjn}
\end{align}
The terms $\varpi_b^r(t)$ and $\varpi_b^v(t)$ in \eqref{rjn} and \eqref{vjn} are observation noises, modeled as additive complex white Gaussian processes with power spectral density $\Pn$. We assume that all noise processes are independent. Let $\br = \{r_b[1:T_{ob}/T]: b\in\stB\}$ and $\bv = \{v_b[1:T_{ob}/T] : b\in\stB\}$ be the collection of observations from all beacons.

Our analysis is based on the received sequences given by \eqref{rjn} and \eqref{vjn}. Treating the agent clock skews as nuisance parameters, there are $4L+2+2N$ unknown parameters $\bp_1, \bp_2, \bv_1, \bv_2, \Omega_1, \Omega_2$ and $\{\beta_b, \Omega_b\}_{b\in\stB}$, where $L$ is the length of the position vector $\bp_1$. We stack the unknown parameters into a vector and denote it as $\bx = [\bp_1^T, \bp_2^T, \bv_1^T, \bv_2^T, \Omega_1, \Omega_2,\{\beta_b, \Omega_b\}_{b\in\stB}]^T$. Let $\hat{\bx}$ be an estimate of $\bx$. To simplify the computations, we use the modified Bayesian CRLB \cite{Trees2007,Moeneclaey1998,Gini1998}, which allows us to first treat the agent clock skews $\beta_1$ and $\beta_2$ as known values, and then taking expectation over all random clock skews. The actual Bayesian CRLB gives a tighter error bound, but has a much more complicated form that unfortunately does not provide additional insights compared to the analysis in this paper. We therefore choose to present the modified Bayesian CRLB instead. We have
\begin{align*}
\EE{}{(\hat{\bx} -  \bx)(\hat{\bx} - \bx)^T} \geq \bF_{\bx}^{-1},
\end{align*}
where $\bF_{\bx}$ is the Fisher information matrix (FIM), which can be shown to be\footnote{The notation $\mathbb{E}_{\condp{y}{x}}$ means taking the expectation over $y$ conditioned on $x$, while $\condp{y}{x}$ is the probability density function of $y$ conditioned on $x$.}
\begin{align}
\bF_{\bx} = \EE{\bbeta}{\EE{\br, \bv|\bx}{-\derdd{\ln\condp{\br, \bv}{\bx,\beta_1,\beta_2}}{\bx}{\bx^T}}}, \label{BFIM}
\end{align}
where $\bbeta = \{\beta_m\}_{m\in\stS\cup\stB}$ are the random clock skews \cite{Moeneclaey1998, Gini1998}.

\subsection{Derivation of $\bF_{\bx}$}
We first map the parameter vector $\bx$ into another parameter vector $\by = [\by_0, \by_1, \cdots, \by_N]^T$, where $\by_0 = [\T_{1,2}, \D_{1,2}, \Omega_1, \Omega_2]$ and $\by_b = [\T_{1,b}, \T_{2,b}, \D_{1,b}, \D_{2,b}, \beta_b, \Omega_b]$ for $b\in\stB$. The FIM for $\bx$ can then be shown to be
\begin{align}
\frac{\Pn}{2} \bF_{\bx} = \EE{\bbeta}{\bJ \bF_{\by} \bJ^T}, \label{Fxy}
\end{align}
where $\bF_{\by}$ is the FIM for $\by$, and $\bJ$ is the Jacobian matrix for the transformation from $\bx$ to $\by$, i.e., $\bJ = \partial \by / \partial \bx$. To derive the expression for $\bJ$, we first obtain $\partial \T_{1,2}/\partial \bp_1 = \bu_{1,2}/c$,  $\partial \D_{1,2}/\partial \bv_1 = -f_0\bu_{1,2}/c$, and
\begin{align*}
\der{\D_{1,2}}{\bp_1} = -\frac{1}{c} \underbrace{\frac{\bI - \bu_{1,2}\bu_{1,2}^T}{\norm{\bp_1 - \bp_2}}(\bv_1 - \bv_2)f_0}_{\triangleq \bw_{1,2}},
\end{align*}
where $\bw_{1,2}$ is the orthogonal projection of $f_0(\bv_1-\bv_2)$ onto a direction normal to $\bu_{1,2}$. Similarly, we have $\partial \D_{j,b}/\partial \bv_j = -f_b\bu_{j,b}/c$ and
\begin{align}
\der{\D_{j,b}}{\bp_j} = -\frac{1}{c}\cdot\underbrace{\frac{\bI - \bu_{j,b}\bu_{j,b}^T}{\norm{\bp_j - \bp_b}}\bv_jf_b}_{\triangleq \bw_{j,b}}, \label{w}
\end{align}
where $\bw_{j,b}$ is orthogonal to $\bu_{j,b}$. Utilizing these facts, it can be shown that $\bJ\in\R^{(4L+2+2N)\times(4+6N)}$ and $\bJ = \partial \by / \partial \bx = [\bJ_0^T, \bJ_1^T, \cdots, \bJ_b^T, \cdots, \bJ_N^T]$, where
\begin{align*}
\bJ_0 & = \left(\der{\by_0}{\bx}\right)^T =
\left[\begin{array}{c:cc}
        \bL_0 & \multicolumn{2}{c}{\bo} \\ \hdashline
        \bo & \bI_2 & \bo
		\end{array}
\right], 
\end{align*}
and
\begin{align*}
\bJ_b = \left(\der{\by_b}{\bx}\right)^T =
\left[\begin{array}{c:ccc}
		\bL_b & \multicolumn{3}{c}{\bo} \\ \hdashline
		\bo & \bo_{2\times2(b+1)} & \bI_2 & \bo
		\end{array}
\right], 
\end{align*}
with
\begin{align}
\bL_0 & = \left[\begin{array}{cccc}
        \frac{1}{c}\bu_{1,2}^T & -\frac{1}{c}\bu_{1,2}^T & \bo & \bo \\
		   -\frac{1}{c}\bw_{1,2}^T & \frac{1}{c}\bw_{1,2}^T & -\frac{f_0}{c}\bu_{1,2}^T & \frac{f_0}{c}\bu_{1,2}^T 
		\end{array}
\right],
\end{align}
and
\begin{align}
\bL_b & = \left[\begin{array}{cccc}
		\frac{1}{c}\bu_{1,b}^T & \bo & \bo & \bo  \\
		\bo & \frac{1}{c}\bu_{2,b}^T & \bo & \bo  \\
		-\frac{1}{c}\bw_{1,b} ^T& \bo & -\frac{f_b}{c}\bu_{1,b}^T & \bo \\
		\bo & -\frac{1}{c}\bw_{2,b}^T & \bo & -\frac{f_b}{c}\bu_{2,b}^T
		\end{array}
\right],
\end{align}
where $\bI_2$ is a $2 \times 2$ identity matrix, $\bo_{a \times b}$ is a $a \times b$ zero matrix, and the notation $\bo$ represents a zero matrix of the appropriate dimensions, which are easily inferred from the context.

Utilizing the well known expression for the complex Gaussian CRLB \cite[p.525]{Kay}, the matrix $\bF_{\by}$ can be obtained as from \eqref{BFIM} and \eqref{Fxy} as
\begin{align}
\bF_{\by} = \sum_b\Real{\sum_l \der{\mu_b^*[l]}{\by}\left(\der{\mu_b[l]}{\by}\right)^T} + \sum_b \Real{\sum_l \der{\theta_b^*[l]}{\by}\left(\der{\theta_b[l]}{\by}\right)^T}. \label{Fy}
\end{align}
Notice that $\partial \mu_b[l] / \partial \by_a = 0$ and $\partial \theta_b[l] / \partial \by_a = 0$ for all $a \neq b$, so that $\bF_{\by}$ is a block matrix with structure
\begin{align*}
\bF_{\by} =
\left[\begin{array}{c:c c c}
        \sum_{b\in\stB} \bH_{b} & \bK_1 & \cdots & \bK_N \\ \hdashline
        \bK_1^T & \bE_{1} & \cdots & \bo \\
        \vdots    & \vdots & \ddots & \vdots \\
        \bK_N^T & \bo & \cdots & \bE_N
      \end{array}
\right],
\end{align*}
where
\begin{align*}
\bE_{b} & = \Real{\sum_l\der{\mu_b^*[l]}{\by_b}\left(\der{\mu_b[l]}{\by_b}\right)^T} + \Real{\sum_l\der{\theta_b^*[l]}{\by_b}\left(\der{\theta_b[l]}{\by_b}\right)^T} , \\
\bH_{b} & = \Real{\sum_l\der{\mu_b^*[l]}{\by_0}\left(\der{\mu_b[l]}{\by_0}\right)^T} + \Real{\sum_l\der{\theta_b^*[l]}{\by_0}\left(\der{\theta_b[l]}{\by_0}\right)^T} , \\
\bK_b & = \Real{\sum_l \der{\mu_b^*[l]}{\by_0}\left(\der{\mu_b[l]}{\by_b}\right)^T} + \Real{\sum_l\der{\theta_b^*[l]}{\by_0}\left(\der{\theta_b[l]}{\by_b}\right)^T} ,
\end{align*}
for $b = 1, \cdots, N$. Substituting $\bJ$ and $\bF_{\by}$ into \eqref{Fxy}, we have
\begin{align}
\frac{\Pn}{2} \bF_{\bx} = \sum_{b\in\stB} \EE{\bbeta}{\bJ_0^T \bH_b \bJ_0 + \bJ_b^T \bE_b \bJ_b + \bJ_b^T\bK_b^T\bJ_0 + \bJ_0^T\bK_b\bJ_b}. \label{FD}
\end{align}

\subsection{Equivalent Fisher Information Matrix}\label{subsect:crlb}
To gain further insights into the structure of $\bF_{\bx}$, we partition the matrices $\bE_b$ such that
\begin{align*}
\bE_b & = \left[
            \begin{array}{c:c}
            \bA_{E,b} & \bB_{E,b} \\ \hdashline
            \bB_{E,b}^T & \bD_{E,b}
            \end{array}
          \right],
\end{align*}
where $\bA_{E,b} \in \R^{4 \times 4}$, $\bB_{E,b} \in\R^{4 \times 2}$, and $\bD_{E,b} \in \R^{2 \times 2}$. Similarly, we have
\begin{align*}
\bH_b =
\left[
    \begin{array}{c:c}
    \bA_{H,b} & \bB_{H,b} \\ \hdashline
    \bB_{H,b}^T & \bD_{H,b}
    \end{array}
\right],
\end{align*}
where $\bA_{H,b}, \bB_{H,b}, \bD_{H,b} \in \R^{2 \times 2}$ and
\begin{align*}
\bK_b & = \left[
            \begin{array}{c:c}
            \bA_{K,b} & \bB_{K,b} \\ \hdashline
            \bC_{K,b} & \bD_{K,b}
            \end{array}
          \right],
\end{align*}
where $\bA_{K,b} \in \R^{2 \times 4}$, $\bB_{K,b} \in\R^{2 \times 2}$, $\bC_{K,b} \in\R^{2 \times 4}$, and $\bD_{K,b} \in \R^{2 \times 2}$. Substituting these matrices into \eqref{FD}, we obtain after some algebra,
\begin{align}
& \frac{\Pn}{2}\bF_{\bx} = \mathbb{E}_{\bbeta}\nonumber\\
& \hspace{-.8cm}\left[
\begin{array}{c:cccc}
\sum_{b\in\stB} \bSigma_b & \sum_{b\in\stB} \left(\bL_0^T\bB_{H,b} + \bL_b^T\bC_{K,b}^T\right) & \bL_0^T\bB_{K,1} + \bL_1^T\bB_{E,1} & \cdots & \bL_0^T\bB_{K,N} + \bL_N^T\bB_{E,N} \\ \hdashline
\sum_{b\in\stB} \left(\bB_{H,b}^T\bL_0 + \bC_{K,b}\bL_b\right) & \sum_{b\in\stB} \bD_{H,b} & \bD_{K,1} & \cdots & \bD_{K,N} \\ 
\bB_{K,1}^T\bL_0 + \bB_{E,1}^T\bL_1 & \bD_{K,1}^T & \bD_{E,1} & \cdots & \bo \\
\vdots & \vdots & \vdots & \ddots & \vdots\\
\bB_{K,N}^T\bL_0 + \bB_{E,N}^T\bL_N & \bD_{K,N}^T & \bo & \cdots & \bD_{E,N}
\end{array}
\right], \label{FIM}
\end{align}
with $\bSigma_b = \bL_b^T\bA_{E,b}\bL_b + \bL_0^T\bA_{H,b}\bL_0 + \bL_0^T\bA_{K,b}\bL_b + \bL_b^T\bA_{K,b}^T\bL_0$. Since we are interested in estimation accuracies for agent locations and velocities, i.e., $\{\bp_j,\bv_j\}_{j=1,2}$, it is sufficient to find its equivalent Fisher information matrix (EFIM) $\bF_{\re,\bs}$ \cite{Trees2007, Shen2010}, which is given by
\begin{align*}
\frac{\Pn}{2}\bF_{\re,\bs} = \sum_{b\in\stB} \EE{\bbeta}{\bSigma_b}- \EE{\bbeta}{\bF_1} \EE{\bbeta}{\bF_2}^{-1} \EE{\bbeta}{\bF_1}^T,
\end{align*}
where $\bF_1$ and $\bF_2$ represent the upper- and lower-right-corner block of $\bF_{\bx}$ in \eqref{FIM}, respectively. Therefore, it follows that
\begin{align}
\frac{\Pn}{2} \bF_{\re,\bs} & = \sum_{b\in\stB} \bQ_b - \left(\sum_{b\in\stB} \bU_b\right)\left(\sum_{b\in\stB} \bV_b\right)^{-1}\left(\sum_{b\in\stB} \bU_b^T\right), \label{EFIM}
\end{align}
where
\begin{align*}
\bQ_b & = \bL_b^T \left(\bbA_{E,b} - \bbB_{E,b}\bbD_{E,b}^{-1}\bbB_{E,b}^T\right) \bL_b
        + \bL_0^T \left(\bbA_{H,b} - \bbB_{K,b}\bbD_{E,b}^{-1}\bbB_{K,b}^T\right) \bL_0 \\
      & + \bL_0^T \left(\bbA_{K,b} - \bbB_{K,b}\bbD_{E,b}^{-1}\bbB_{E,b}^T\right) \bL_b
        + \bL_b^T \left(\bbA_{K,b}^T - \bbB_{E,b}\bbD_{E,b}^{-1}\bbB_{K,b}^T\right) \bL_0, \\
\bU_b & = \bL_0^T \left(\bbB_{H,b} - \bbB_{K,b}\bbD_{E,b}^{-1}\bbD_{K,b}^T\right)
        + \bL_b^T \left(\bbC_{K,b}^T - \bbB_{E,b}\bbD_{E,b}^{-1}\bbD_{K,b}^T\right), \\
\bV_b & = \bbD_{H,b} - \bbD_{K,b}\bbD_{E,b}^{-1}\bbD_{K,b}^T,
\end{align*}
with $\bar{\bA} = \EE{\bbeta}{\bA}$ for any matrix $\bA$.

To facilitate further analysis of the EFIM in \eqref{EFIM}, we define the signal energy $\Pb$, the root-mean-square (RMS) bandwidth $\Wb$, and the RMS integration time $\Tb$ for the signal $g_b(t)$ as \cite{Auger2008},
\begin{align*}
\Pb = \int \abs{g_b(t)}^2 \dd t, \quad 
\Wb = \left[\frac{\int \abs{f G_b(f)}^2 \dd f}{\int \abs{G_b(f)}^2 \dd f}\right]^{\frac{1}{2}}, \quad
\Tb = \left[\frac{\int \abs{t g_b(t)}^2 \dd t}{\int \abs{g_b(t)}^2 \dd t}\right]^{\frac{1}{2}}.
\end{align*}
Without loss of generality, we assume that the signal $g_b(t)$ has zero centroid in time and frequency, hence $\Wb$ and $\Tb$ characterize the signal's energy dispersion around its centroid in time and frequency, respectively.
In the following, we make various assumptions and approximations, which hold in most practical applications. We use $a \ll b$ to mean that $a/b$ can be approximated by $0$.
\begin{assumption} \label{assumption}\
\begin{enumerate}[(i)]
\item The clock skew standard deviations $\sigma_{\beta_m} < 1$ for all $m \in\stS\cup\stB$.
\item {For every beacon $b\in\stB$, the RMS bandwidth $\Wb$ is much smaller than the nominal carrier frequency $f_b$ with $\Wb \ll f_b$.}
\item {There exists $\epsilon >0$ and a measurable set with probability at least $1-\epsilon$ so that $\T_{1,2}\ll \Tb/3$, $\T_{j,b} \ll \Tb/3$, and $\Omega_j \ll \beta_j\Tb/3$, for $j=1,2$ and for every beacon $b\in\stB$. Futhermore, $\epsilon$ can be chosen sufficiently small so that all expectations can be approximated by taking expectations over this set.}
\end{enumerate}
\end{assumption}
For each $b\in\stB$, let $\SNRb = \Pb/\Pn$ be the effective output signal-to-noise ratio of the received signal from $b$. As shown in \cite{Stein1981}, the effective output $\SNRb$ depends on the input SNR and the bandwidth-time product $\mathsf{W}_b\mathsf{T}_b$. Let
\begin{align}
&\lambda_b = \frac{8\pi^2\Wb^2\SNRb}{c^2}, \textrm{ and } \epsilon_b = \frac{8\pi^2\Tb^2\SNRb}{c^2}. \label{chi-omega}
\end{align}
We have the following theorem, whose proof is given in Appendix \ref{appdx-crlb}.
\begin{theorem}\label{theorem:crlb} \
Suppose that Assumption \ref{assumption} holds. Let $\bphi_b = [\bu_{1,b}^T, -\bu_{2,b}^T]^T$, $\bphi_{s} = [\bu_{1,2}^T, -\bu_{1,2}^T]^T$, $\brho_b = [\bw_{1,b}^T, -\bw_{2,b}^T]^T$, and $\brho_{s} = [\bw_{1,2}^T, -\bw_{1,2}^T]^T$.  Treating the agent clock skews as nuisance parameters, the EFIM for estimating the agents' locations and velocities at agent $S_1$, denoted as $\bF_{\re,\bs}$, is approximately given by
\begin{align}
    \bF_{\re,\bs} & = \frac{1}{2}\sum_{b} 
        \begin{bmatrix}
        \lambda_b(\bPhi_b -\bXi)
        +\epsilon_b(1-\sigma_{\beta_b}^2)^{-1}\bRho_b & \epsilon_b(1-\sigma_{\beta_b}^2)^{-1}\bGamma_b \\
        \epsilon_b(1-\sigma_{\beta_b}^2)^{-1}\bGamma_b^T & \epsilon_b(1-\sigma_{\beta_b}^2)^{-1}\bPi_b
        \end{bmatrix}, \label{EFIMr}
\end{align}
where we let $\bar{\lambda}_b = \frac{\lambda_b}{\sum_{b'} \lambda_{b'}}$, $\sigma_s = 2[1+(\sigma_{\beta_1}^2-3\sigma_{\beta_1}^2\sigma_{\beta_2}^2)/(1-3\sigma_{\beta_2}^2+\sigma_{\beta_1}^2\sigma_{\beta_2}^2)]$, and
\begin{align*}
	\bXi & = \sigma_s \sum_b \bar{\lambda}_b(\bphi_b - \bphi_s)\sum_b \bar{\lambda}_b(\bphi_b - \bphi_s)^T, \\					 
	\bPhi_b & = (\bphi_b - \bphi_s)(\bphi_b - \bphi_s)^T, \\
  \bRho_b & = (\brho_b - \brho_s)(\brho_b - \brho_s)^T, \\  
   \bPi_b & = (f_b\bphi_b - f_0\bphi_s)(f_b\bphi_b - f_0\bphi_s)^T, \\
    \bGamma_b & = (\brho_b-\brho_s)(f_b\bphi_b - f_0\bphi_s)^T.
    \end{align*}
\end{theorem}

Theorem \ref{theorem:crlb} provides an approximate lower bound for the location and velocity estimation errors. However, since we have used the modified Bayesian CRLB, this bound is not tight. Nevertheless, the information bound in Theorem \ref{theorem:crlb} gives us various insights into the problem of location and velocity estimation, which we discuss below.
\begin{enumerate}[1)]
\item We see from \eqref{EFIMr} that the EFIM consists of various ``information matrix'' components, which we describe in the following.
\begin{itemize}
	\item The matrix $\bPhi_b$ can be interpreted as the ranging direction matrix (RDM) associated with the directions $\bu_{1,b} - \bu_{1,2}$ and $\bu_{2,b} - \bu_{1,2}$. This is similar to the RDM introduced in \cite{Shen2010}, where the EFIM is derived in the case where a single agent localizes with the aid of synchronized anchors so that $\bu_{1,2} = \bo$. Similar to \cite{Shen2010}, we can interpret $\lambda_b$ as the \textit{ranging information intensity}, which is a constant that depends only on beacon characteristics like SNR. We note that the beacon clock skews do not affect the ranging information intensity or the RDM, which is intuitively correct as TDOA ranging is not affected by beacon clock skews.
	\item Let $\bu^\perp$ be the unit vector orthogonal to $\bu$. The vectors $\brho_b$ and $\brho_s$ contain Doppler shift information in the directions $\bu_{1,b}^\perp$ and $\bu_{1,2}^\perp$ respectively. Therefore, the matrix $\bRho_b$ can be interpreted as the relative Doppler information matrix associated with the directions $\bu_{1,b}^\perp$, $\bu_{2,b}^\perp$ and $\bu_{1,2}^\perp$. This is intuitively appealing as all location information in the directions $\bu_{1,b}$, $\bu_{2,b}$, and $\bu_{1,2}$ have already been captured in $\bPhi_b$ so that $\bRho_b$ contains additional information in directions orthogonal to these. Moreover, since Doppler shift is affected by beacon clock skews, we have a factor of $(1-\sigma_{\beta_b}^2)^{-1}$ multiplied to $\bRho_b$ in \eqref{EFIMr}. We can also interpret $\epsilon_b$ as the \textit{Doppler information intensity}. 
	\item The term $f_b\bphi_b$ is the rate of change of Doppler shift in the direction $\bu_{j,b}$ w.r.t. $\bv_j$. Therefore, the matrix $\bPi_b$ contains information associated with how fast the Doppler shift along the directions $\bu_{j,b}$ is changing w.r.t. that along $\bu_{1,2}$. This is mainly useful for velocity estimation, and so do not appear in the CRLB derived in \cite{Shen2010, Yeredor2011}. Beacon clock skews affect the information contained in $\bPi_b$ and appears in the multiplicative factor $(1-\sigma_{\beta_b}^2)^{-1}$ in \eqref{EFIMr}.
	\item The term $\bXi$ contains the weighted average ranging information from all beacons, with the weight of beacon $b$ being $\lambda_b$ normalized by the sum of all ranging information intensities. Since $S_2$ transmits all its received signals from the beacons to $S_1$, we can interpret $\bXi$ as the collective effect of agent clock asychronism on the information transmitted from $S_2$. 
\end{itemize}

\item From \eqref{EFIMr}, it is clear that the EFIM for agent locations and velocities depend on neither the value of beacon clock offsets nor the value of agent clocks offsets. This suggests that there exists estimation algorithms that can eliminate \textit{both} beacon and agent clock offsets. It can be shown that the TDOA procedure cancels out the beacon clock offsets. We will show in Section \ref{Section:Algorithm} that the DTDOA method allows us to cancel out agent clock offsets as well.

\item Although the size of clock offsets do not affect the EFIM, we observe that there is loss of information whenever agent clocks are not synchronized. Consider the case where the two agents and beacons are static, so that $\brho_b = \bo$ and $\brho_s = \bo$. The EFIM in \eqref{EFIMr} for agent locations reduces to
\begin{align}
\bF_{\re, static} = \frac{1}{2}\sum_{b}\lambda_b\left(\bPhi_b-\bXi\right). \label{Fstatic}
\end{align}
On the other hand, if agents and beacons are static and synchronized, it can be shown that the EFIM is given by
\begin{align}
\bF_{\re, syn} = \sum_{b}\lambda_b\left(\bPhi_b + \begin{bmatrix}\bu_{1,b}\bu_{1,2}^T+\bu_{1,2}\bu_{1,b}^T & \bu_{1,b}(\bu_{2,b}-\bu_{1,2})^T \\ (\bu_{2,b}-\bu_{1,2})\bu_{1,b}^T & \bo \end{bmatrix}\right). \label{Fsyn}
\end{align}
Therefore, the information loss due to agent clock offsets is given by $\bF_{\re,syn}-\bF_{\re, static}$ which is non-zero. It can be seen that the information loss does not depend on the value of agent clock offsets, suggesting that clock offsets can be eliminated but at the price of a constant information loss.

\item As can be seen from \eqref{EFIMr}, the EFIM depends on direction vectors $\bu_{j,b}$ and $\bu_{1,2}$ which are determined only by the relative positions of agents and beacons. To analyze the effect of beacon-agent geometry on the localization performance, we consider the case \eqref{Fstatic} in two dimensional space with static agents and beacons uniformly distributed on a circle centered at $S_1$. Letting the angle from beacon $b$ to $S_j$ be $\kappa_{j,b}$, we obtain $\bu_{j,b}=[\cos\kappa_{j,b}, \sin\kappa_{j,b}]^T$. As the radius of the circle increases, we have in the limit of infinite radius, $\kappa_{1,b}=\kappa_{2,b}$ so that
\begin{align*}
\lim_{\norm{\bp_b-\bp_1}\rightarrow \infty} \bu_{1,b} = \lim_{\norm{\bp_b-\bp_2}\rightarrow \infty} \bu_{2,b} \triangleq \breve{\bu}_b.
\end{align*}
The EFIM in \eqref{Fstatic} then reduces to
\begin{align*}
\bF_{\re,static} = \sum_b \frac{\lambda_b}{2} \begin{bmatrix} \breve{\bPhi}_b & -\breve{\bPhi}_b \\ -\breve{\bPhi}_b & \breve{\bPhi}_b\end{bmatrix},
\end{align*}
with $\breve{\bPhi}_b = (\breve{\bu}_b+\breve{\bu}_{1,2})(\breve{\bu}_b+\breve{\bu}_{1,2})^T - \sigma_s\sum_{b}\bar{\lambda}_b\breve{\bu}_b\sum_{b}\bar{\lambda}_b\breve{\bu}_b^T - \sigma_s\sum_{b}\bar{\lambda}_b\breve{\bu}_b\bu_{1,2}^T - \sigma_s\bu_{1,2}\sum_{b}\bar{\lambda}_b\breve{\bu}_b^T - \sigma_s\bu_{1,2}\bu_{1,2}^T$. 
It is clear that the corresponding EFIM $\bF_{\re, s}$ is rank-deficient, which implies that large estimation errors will be introduced when beacons are too far away from the agents.

\item Suppose that agent $S_2$ has known location and velocity. We then have a simpler expression for the EFIM. We have $\bphi_b = \bu_{1,b}$, $\bphi_s=\bu_{1,2}$, $\brho_b=\bw_{1,b}$ and $\brho_s=\bw_{1,2}$, with
$ \bPhi_b = (\bu_{1,b}-\bu_{1,2})(\bu_{1,b}-\bu_{1,2})^T, $ 
$ \bRho_b = (\bw_{1,b}-\bw_{1,2})(\bw_{1,b}-\bw_{1,2})^T, $
$ \bPi_b = (f_b\bu_{1,b}-f_0\bu_{1,2})(f_b\bu_{1,b}-f_0\bu_{1,2})^T, $ and
$ \bGamma_b = (\bw_{1,b}-\bw_{1,2})(f_b\bu_{1,b}-f_0\bu_{1,2})^T.$
In this case, $S_2$ acts almost like an anchor, except that it does not transmit its own independent signal. Therefore the amount of information available to $S_1$ is less than that of a true anchor.
\end{enumerate}

\section{Conclusions}\label{sect:con}
In this paper, we have considered the problem of localizing two agents using signals of opportunity from beacons with known locations. We assume that all beacons and agents have free-running oscillators that have unknown clock skews and offsets. Each agent performs self-localization by using TDOA and FDOA measurements between the two agents. We analyze the biases introduced by asynchronous clocks into the received signals and obtained closed-form expressions for TDOA and FDOA estimates based on these distorted signals. We derive the equivalent Fisher information matrix for the modified Bayesian CRLB using the received signal waveforms. 

In this paper, we have assumed for simplicity that agents forward received signals from the beacons to each other. This may be impractical or may result in high communication costs. Future research includes investigating the more practical scenario where agent information exchanges are limited in bandwidth and exchange frequency. We have also limited our investigations to static beacons in this paper. The use of mobile beacons like unmanned aerial vehicles and non-GPS satellites is of practical interest.

\appendices
\section{Proof of Proposition \ref{proposition:signals}} \label{appdx-signal}
\begin{IEEEproof}
\eqref{it:gen} The baseband signal is generated at beacon $b$. The frequency scaling and time delay are introduced due to its local oscillator. Since $\tilde{g}_b(t) = g_b(\beta_bt + \Omega_b)$, the result follows directly.

\eqref{it:up} The carrier wave generated at beacon $b$ is $\exp\{\ri f_b(\beta_bt + \Omega_b)\}$, and the emitted signal $\tilde{u}_b(t)$ is obtained by multiplying the baseband signal $\tilde{g}_b(t)$ with the carrier wave. This operation introduces a frequency and phase shift into the baseband signal, and the result follows easily.

\eqref{it:prop} The received signal $\tilde{r}_{j,b}(t)$ is obtained by the convolution of the emitted signal $\tilde{u}_b(t)$ and the channel impulse $h_{j,b}(t)$, and we have $\tilde{r}_{j,b}(t) = \alpha_{j,b}\tilde{u}_b(t - \tau_{j,b}(t)) = \alpha_{j,b}\tilde{u}_b(\gamma_{j,b}t - \T_{j,b})$, from which the positive frequency part $\tilde{R}_{j,b}(\xi)$ follows easily.

\eqref{it:down} The down-conversion of the received signal $\tilde{r}_{j,b}(t)$ to the baseband is achieved by multiplying $\tilde{r}_{j,b}(t)$ with a locally generated carrier wave and passing it through a low-pass filter. The carrier wave generated by $S_j$ is given by $\exp\{-\ri f_b(\beta_jt + \Omega_j)\}$, and hence we have the result in \eqref{it:down}.

\eqref{it:sample} Consider the sampling procedure at $S_j$. The sampling impulse train is given by $\sum_{l = -\infty}^{+\infty} \delta(\beta_jt - lT)$. Denoting $t_0$ as the starting point of sampling, and $T_{ob}$ as the observation interval, we obtain the signal sample sequence $r_{j,b}[l]= r_{j,b}(lT/\beta_j)$ for $l = t_0/T : (t_0+T_{ob})/T$. We denote $\bar{r}_{j,b}(t) = \sum_{l} r_{j,b}[l] \psi_l(t)$, where $\psi_l(t) = 1$ for the sample sequence. In general, $\{\psi_l(t)\}$ constitutes a complete orthonormal basis in the space of band-limited $\mathcal{L}_2$ functions, for example, $\psi_l(t) = \sin[\pi (t/T - l)]/[\pi (t/T - l)]\}_{l = -\infty}^{+\infty}$. The Fourier transform $\bar{R}_j(\xi)$ is,
\begin{align*}
\bar{R}_{j,b}(\xi) & = \sum_{l}\int r_{j,b}[l]\psi_l(t) \re^{-\ri \xi t} \dd t \nonumber\\
           & = \rec{T\xi} \sum_l r_{j,b}[l]\re^{- \ri \xi lT} \ = R_{j,b}\left(\beta_j\xi\right),
\end{align*}
where $\rec{\cdot}$ denotes the standard rectangular function and acts as an ideal low-pass filter. Therefore, the sampled signal converges in $\mathcal{L}_2$ to a time-scaled version of the original signal.
\end{IEEEproof}

\section{Proof of Theorem \ref{theorem:signals}} \label{appdx-rjvj}
\begin{IEEEproof}
We first show the signal model $R_{1,b}(\xi)$ from beacon $b$ to agent $S_1$. Using Proposition \ref{proposition:signals}, the received signal at $S_1$ from beacon $b$ can be obtained by substituting Proposition \ref{proposition:signals} \eqref{it:up} and \eqref{it:prop} into \eqref{it:down}, and we have
\begin{align}
\tilde{R}_{1,b}(\xi) & = \frac{\alpha_{1,b}}{\gamma_{1,b}}\tilde{U}_b\left(\frac{\xi}{\gamma_{1,b}}\right)\re^{-\ri\xi\frac{\T_{1,b}}{\gamma_{1,b}}} \nonumber\\
& = \frac{\alpha_{1,b}}{\gamma_{1,b}}\tilde{G}_b\left(\frac{\xi}{\gamma_{1,b}} - f_b\beta_b\right)\re^{-\ri \xi\frac{\T_{1,b}}{\gamma_{1,b}}}\re^{\ri f_b\Omega_b}, \label{trj}
\end{align}
and hence
\begin{align}
R_{1,b}(\xi) & = \tilde{R}_{1,b}(\xi + f_b\beta_1)\re^{-\ri f_b \Omega_1} \nonumber\\
& = \frac{\alpha_{1,b}}{\gamma_{1,b}}\tilde{G}_b\left(\frac{\xi - \Upsilon_{1,b}}{\gamma_{1,b}}\right)\re^{-\ri \xi \frac{\T_{1,b}}{\gamma_{1,b}}} \re^{\ri \frac{\Upsilon_{1,b}\T_{1,b}}{\gamma_{1,b}}} \re^{\ri \varphi_{1,b}} \nonumber\\
& \propto \tilde{G}_b\left(\frac{\xi - \Upsilon_{1,b}}{\gamma_{1,b}}\right)\re^{-\ri \xi \frac{\T_{1,b}}{\gamma_{1,b}}},
\end{align}
where $\varphi_{1,b} = f_b(\Omega_b - \Omega_1 - \beta_b\T_{1,b})$, $\Upsilon_{1,b} = \D_{1,b}\beta_b + f_b(\beta_b - \beta_1)$. Notice that signals used for the estimation of TDOA and FDOA should be received from the same beacon within the same interval, we assume two agents make an agreement that they start to collect incoming signals at their local time $t_0$. This agreement can be achieved by a prior handshake procedure between two agents, and we assume $t_0 = 0$ without loss of generality. Since two agents are asynchronous, the received signal at $S_1$ will actually start at the standard time of $-\Omega_1/\beta_1$ and the collected signal is $r_{1,b}(t + \Omega_1/\beta_1)$. Therefore, we further have
\begin{align}
R_{1,b}(\xi) & \propto \tilde{G}_b\left(\frac{\xi - \Upsilon_{1,b}}{\gamma_{1,b}}\right)\re^{-\ri \xi \frac{\T_{1,b} - \Omega_1/\beta_1}{\gamma_{1,b}}}. \label{rj}
\end{align}
Substituting Proposition \ref{proposition:signals}\eqref{it:gen} and \ref{proposition:signals}\eqref{it:sample} into \eqref{rj}, the result \eqref{r1} follows easily.

Similarly, we show the signal model $V_{1,b}(\xi)$ from $S_2$ to $S_1$. Two agents communicate with each other by transmitting narrowband signals with nominal carrier frequency $f_0$. In order to forward its received signal to $S_1$, $S_2$ shifts the frequency of $r_{2,b}(t)$ to $f_0$ using frequency mixer, and obtains the emitted signal, denoted as $\tilde{u}_{2,b}(t)$. Using Proposition \ref{proposition:signals}\eqref{it:up}, $\tilde{u}_{2,b}(t)$ has positive frequency part,
\begin{align}
\tilde{U}_{2,b}(\xi) & = R_{2,b}(\xi - f_0\beta_2)\re^{\ri 2f_0\Omega_2}. \label{tuk}
\end{align}
Similar to that in \eqref{rj}, the collected signal at $S_2$ starts at $-\Omega_2/\beta_2$, and it follows that
\begin{align}
R_{2,b}(\xi) \propto \tilde{G}_b\left(\frac{\xi - \Upsilon_{2,b}}{\gamma_{2,b}}\right)\re^{-\ri \xi \frac{\T_{2,b} - \Omega_2/\beta_2}{\gamma_{2,b}}}.
\end{align}
The signal $\tilde{u}_{2,b}(t)$ is then forwarded to $S_1$ through the channel $h_{1,2}(t)$. Using Proposition \ref{proposition:signals}\eqref{it:prop}, the received signal at $S_1$, denoted as $\tilde{v}_{1,b}(t)$, is given by
\begin{align}
\tilde{V}_{1,b}(\xi) & = \frac{\alpha_{1,2}}{\gamma_{1,2}}\tilde{U}_{2,b}\left(\frac{\xi}{\gamma_{1,2}}\right)\re^{-\ri \xi \frac{\T_{1,2}}{\gamma_{1,2}}}. \label{tvj}
\end{align}
The received signal $\tilde{v}_{1,b}(t)$ is then down-converted to baseband using a locally generated carrier wave. From Proposition \ref{proposition:signals}\eqref{it:down}, the down-converted signal at $S_1$, denoted as $v_{1,b}(t)$, is given by
\begin{align}
V_{1,b}(\xi) = \tilde{V}_{1,b}(\xi + f_0\beta_1)\exp\{-\ri f_0\Omega_1\}. \label{vj}
\end{align}
Therefore, substituting \eqref{tuk}--\eqref{tvj} and Proposition \ref{proposition:signals}\eqref{it:gen} into \eqref{vj}, it can be shown that
\begin{align}
V_{1,b}(\xi) & = \frac{\alpha_{1,2}\alpha_{2,b}}{\gamma_{1,2}\gamma_{2,b}} \tilde{G}_b\left(\frac{\xi - \Psi_{1,b}}{\gamma_{1,2}\gamma_{2,b}}\right) \re^{-\ri \xi \frac{\tilde{\Lambda}_{1,b}}{\gamma_{1,2}\gamma_{2,b}}}\re^{\ri \frac{\Psi_{1,b}\tilde{\Lambda}_{1,b}}{\gamma_{1,2}\gamma_{2,b}}}\re^{\ri \varphi_1} \nonumber\\
& \propto G_b\left(\frac{\xi - \Psi_{1,b}}{\beta_b\gamma_{1,2}\gamma_{2,b}}\right) \re^{-\ri \xi \frac{\beta_b\tilde{\Lambda}_{1,b} - \Omega_b}{\beta_b\gamma_{1,2}\gamma_{2,b}}},
\end{align}
where $\varphi_1 = \varphi_{2,b} + \varphi_{1,2} - \Upsilon_{2,b}(\T_{1,2} - \Omega_2/\beta_2)$, $\Psi_{1,b} = \left[\D_{2,b}\beta_b + f_b(\beta_b - \beta_2)\right]\gamma_{1,2} + [\D_{1,2}\beta_2 + f_0(\beta_2 - \beta_1)]$, and $\tilde{\Lambda}_{1,b} = \gamma_{2,b}\T_{1,2} + \T_{2,b} - \Omega_2/\beta_2$. Finally from Proposition \ref{proposition:signals}\eqref{it:sample}, we have the result in \eqref{v1}.
\end{IEEEproof}

\section{Proof of Theorem \ref{theorem:crlb}} \label{appdx-crlb}
\begin{IEEEproof}
We first derive expressions for submatrices of $\{\bE_b, \bH_b, \bK_b\}_{b=1}^{N}$. We summarize the results in the following Lemma \ref{lemma:sub}, whose proof is given in Appendix \ref{appdx-sub}.
\begin{lemma}\label{lemma:sub}\
\begin{enumerate}[(i)]
\item \label{lemma:subD}
    Suppose Assumption \ref{assumption} holds, the block matrix $\bV_b$ in \eqref{EFIM} is given by
    \begin{align}
  \bV_b & \approx \frac{c^2\Pn\lambda_b}{4}\begin{bmatrix}
				  1/(1-\sigma_{\beta_1}^2) &  -1/[(1-\sigma_{\beta_1}^2)(1-\sigma_{\beta_2}^2)]\\
					-1/[(1-\sigma_{\beta_1}^2)(1-\sigma_{\beta_2}^2)] & \sigma_{\beta_2}^2/[(1-\sigma_{\beta_2}^2)^2(1-2\sigma_{\beta_2}^2)]
					\end{bmatrix}. \label{Vb}
    \end{align}
\item \label{lemma:subE}
    Suppose Assumption \ref{assumption} holds, we have
    \begin{align}
    & \bbA_{E,b} - \bbB_{E,b}\bbD_{E,b}^{-1}\bbB_{E,b}^T \approx \frac{c^2\Pn}{4}
		\begin{bmatrix}
     \lambda_b &  -\lambda_b &        0 &               0\\
     -\lambda_b&       \lambda_b &                 0 &               0 \\
     0 &   0 & \epsilon_b/(1-\sigma_{\beta_b}^2) &                    -\epsilon_b/(1-\sigma_{\beta_b}^2) \\
     0 &   0 &                 -\epsilon_b/(1-\sigma_{\beta_b}^2)     & \epsilon_b/(1-\sigma_{\beta_b}^2)
	  \end{bmatrix}, \label{lemma:Eb} \\
    & \bbA_{H,b} - \bbB_{K,b}\bbD_{E,b}^{-1}\bbB_{K,b}^T \approx \frac{c^2\Pn}{4} 
		\begin{bmatrix}
	  \lambda_b &   0 \\
		0 & \epsilon_b/(1-\sigma_{\beta_b}^2)
    \end{bmatrix}, \label{lemma:E0} \\
    & \bbA_{K,b} - \bbB_{K,b}\bbD_{E,b}^{-1}\bbB_{E,b}^T \approx \frac{c^2\Pn}{4} 
		\begin{bmatrix}
	  -\lambda_b &  \lambda_b &   0 &         0 \\
		0 & 0 & -\epsilon_b/(1-\sigma_{\beta_b}^2) & \epsilon_b/(1-\sigma_{\beta_b}^2)
		\end{bmatrix}. \label{lemma:Kb}
    \end{align}
\end{enumerate}
\end{lemma}
Next, we prove Theorem \ref{theorem:crlb} using the above results. It can be seen from \eqref{EFIM} that the EFIM for agent location and velocity is given by
\begin{align}
\bF_{\re,\bs} = \frac{2}{\Pn}\sum_{b\in\stB} \bQ_b - \left(\frac{2}{c\Pn}\sum_{b\in\stB} \bU_b\right)\left(\frac{2}{c^2\Pn}\sum_{b\in\stB} \bV_b\right)^{-1}\left(\frac{2}{c\Pn}\sum_{b\in\stB} \bU_b^T\right) , \label{appdx-EFIMr}
\end{align}
which holds if and only if the matrix $\sum_b \bV_b$ is invertible. For the first term in \eqref{appdx-EFIMr}, by substituting \eqref{lemma:Eb}--\eqref{lemma:Kb} into \eqref{EFIM}, we can obtain
\begin{align}
\bQ_b = \frac{\Pn}{4}\begin{bmatrix}
        \lambda_b\bPhi_b + \epsilon_b\bRho_b/(1-\sigma_{\beta_b}^2) & \epsilon_b\bGamma_b/(1-\sigma_{\beta_b}^2) \\
        \epsilon_b\bGamma_b^T/(1-\sigma_{\beta_b}^2) & \epsilon_b\bPi_b/(1-\sigma_{\beta_b}^2)
        \end{bmatrix}, \label{Qb}
\end{align}
with $\bPhi_b$, $\bPi_b$, $\bRho_b$, and $\bGamma_b$ given in \eqref{EFIMr}, where we have utilized the fact that $\EE{}{\beta_m} = 1$ and $\EE{}{1/\beta_m} = 1/(1 - \sigma_{\beta_m}^2)$ for $m \in \stS\cup\stB$ when taking expectations. Similarly, it can be shown that
\begin{align*}
\bbB_{H,b}  - \bbB_{K,b}\bbD_{E,b}^{-1}\bbD_{K,b}^T & \approx \frac{c^2\Pn\lambda_b}{4}
\begin{bmatrix}
	1/(1-\sigma_{\beta_1}^2) & -1/(1-\sigma_{\beta_2}^2) \\ 
	0 & 0
\end{bmatrix}\blue{,} \\
\bbC_{K,b}^T - \bbB_{E,b}\bbD_{E,b}^{-1}\bbD_{K,b}^T & 
\approx \frac{c^2\Pn\lambda_b}{4}
\begin{bmatrix}
	-1/(1-\sigma_{\beta_1}^2) &     1/(1-\sigma_{\beta_2}^2)\\
	1/(1-\sigma_{\beta_1}^2) &     -1/(1-\sigma_{\beta_2}^2)\\
	0 & 0\\
	0 & 0
\end{bmatrix}.
\end{align*}
Substituting these two terms into \eqref{EFIM} and utilizing \eqref{Vb} and Assumption \ref{assumption}, we obtain
\begin{align}
\left(\frac{2}{c\Pn}\sum_{b\in\stB} \bU_b\right)\left(\frac{2}{c^2\Pn}\sum_b \bV_b\right)^{-1}\left(\frac{2}{c\Pn}\sum_{b\in\stB} \bU_b^T\right) \approx
\sum_b\frac{\lambda_b}{2}\begin{bmatrix} \bXi & \bo \\ \bo & \bo \end{bmatrix}, \label{UV}
\end{align}
with $\bXi$ in \eqref{EFIMr}.
\end{IEEEproof}

\section{Proof of Lemma \ref{lemma:sub}}\label{appdx-sub}
\begin{IEEEproof}
To begin with, the expressions for $\{\bE_b, \bH_b, \bK_b\}_{b=1}^{N}$ are derived using their definitions from \eqref{Fy}. For the first term in \eqref{Fy}, denoting $\bmeta_{\mu} = [\Delta_{1,b}, \Upsilon_{1,b}]^T$ and utilizing the chain rule of derivative, we have
\begin{align}
\Real{\sum_l \der{\mu_b^*[l]}{\by_m}\left(\der{\mu_b[l]}{\by_n}\right)^T}
& = \der{\bmeta_{\mu}}{\by_m} \underbrace{\Real{\sum_l \der{\mu_b^*[l]}{\bmeta_{\mu}}\der{\mu_b[l]}{\bmeta_{\mu}^T}}}_{\triangleq \bH_{\mu}}\der{\bmeta_{\mu}}{\by_n^T}, \label{first}
\end{align}
where $m,n\in\{0, b\}$, and
\begin{align*}
\der{\mu_b[l]}{\bmeta_{\mu}} = \re^{-\ri \Upsilon_{1,b} t/\beta_1}
    \begin{bmatrix}
    \partial g_b(\beta_b/\beta_1 t - \Delta_{1,b}) / \partial \Delta_{1,b} \\
    -\ri t/\beta_1 g_b(\beta_b/\beta_1 t - \Delta_{1,b})
    \end{bmatrix}_{t = lT}.
\end{align*}
Using definitions of $\Pb$, $\Tb$ and $\Wb$, it can be shown that \cite{Auger2008}
\begin{align*}
\sum_l (2\pi l T)^2 {g_b^*(\alpha t - \tau)}|_{t = lT}{g_b(\alpha t - \tau)}|_{t=lT} & =  \frac{1}{\alpha^3}4\pi^2\Pb\Tb^2, \\
\sum_l \der{{g_b^*(\alpha t - \tau)}|_{t = lT}}{\tau} \der{{g_b(\alpha t - \tau)}|_{t = lT}}{\tau} & = \frac{1}{\alpha} 4\pi^2\Pb\Wb^2,
\end{align*}
where $T$ represents the sampling interval, and we have used the approximation of Riemann sum with $\lim_{T \rightarrow 0} \sum_l g_b(lT)T \approx \int g_b(t)\dd t$. It hence follows that
\begin{align*}
\bH_{\mu} & = 4\pi^2\Pb \beta_1
    \begin{bmatrix}
    \Wb^2/\beta_b & 0 \\
    0 &\Tb^2/\beta_b^3
    \end{bmatrix},
\end{align*}
where we have assumed that $\gamma_{1,2} \approx 1$ and $\gamma_{j,b} \approx 1$ for $j = 1, 2$. Moreover, recall that $\Delta_{1,b} = \T_{1,b}\beta_b - \Omega_1\beta_b/\beta_1 - \Omega_b$ and $\Upsilon_{1,b} = f_b(\gamma_{1,b}\beta_b-\beta_1)$, we have
\begin{align}
\der{\bmeta_{\mu}}{\by_0} & =
    \begin{bmatrix}
    0 & 0 & -\beta_b/\beta_1 & 0 \\
    0 & 0 & 0 & 0
    \end{bmatrix}^T, \label{mu0}\\
\der{\bmeta_{\mu}}{\by_b} & =
    \begin{bmatrix}
    \beta_b & 0 & 0 & 0 & \T_{1,b}-\Omega_1/\beta_1 & -1 \\
    0 & 0 & \beta_b & 0 & f_b\gamma_{1,b} & 0
    \end{bmatrix}^T. \label{mub}
\end{align}
Substituting $\bH_{\mu}$, \eqref{mu0}, and \eqref{mub} into \eqref{first}, we can obtain the expression for the first term in \eqref{Fy}.

Similarly, for the second term in \eqref{Fy}, denoting $\bmeta_{\theta} = [\Lambda_{1,b}, \Psi_{1,b}]^T$ and utilizing the chain rule of derivative, we have
\begin{align}
\Real{\sum_l \der{\theta_b^*[l]}{\by_m}\left(\der{\theta_b[l]}{\by_n}\right)^T}
& = \der{\bmeta_{\theta}}{\by_m} \underbrace{\Real{\sum_l \der{\theta_b^*[l]}{\bmeta_{\theta}}\der{\theta_b[l]}{\bmeta_{\theta}^T}}}_{\triangleq \bH_{\theta}}\der{\bmeta_{\theta}}{\by_n^T}, \label{second}
\end{align}
where $m,n\in\{0, b\}$. It can be shown that $\bH_{\theta} = \bH_{\mu}$, and recall that $\Lambda_{1,b} = \T_{2,b}\beta_b+\T_{1,2}\gamma_{2,b}\beta_b-\Omega_2\beta_b/\beta_2-\Omega_b$ and $\Psi_{1,b}=f_b\gamma_{1,2}(\gamma_{2,b}\beta_b-\beta_2) + f_0(\gamma_{1,2}\beta_2 - \beta_1)$, we have
\begin{align}
\der{\bmeta_{\theta}}{\by_0} & =
    \begin{bmatrix}
    \gamma_{2,b}\beta_b & 0 & 0 & -\beta_b/\beta_2 \\
    0 & \beta_b\gamma_{2,b}f_b/f_0 + \beta_2(1-f_b/f_0) & 0 & 0
    \end{bmatrix}^T, \label{theta0}\\
\der{\bmeta_{\theta}}{\by_b} & =
    \begin{bmatrix}
    0 & \beta_b & 0 & \T_{1,2}\beta_b/f_b & \T_{2,b}+\T_{1,2}\gamma_{2,b}-\Omega_2/\beta_2 & -1 \\
    0 & 0 & 0 & \gamma_{1,2}\beta_b & f_b\gamma_{1,2}\gamma_{2,b} & 0
    \end{bmatrix}^T. \label{thetab}
\end{align}
Substituting $\bH_{\theta}$, \eqref{theta0}, and \eqref{thetab} into \eqref{second}, we can obtain the expression for the second term in \eqref{Fy}.

Therefore, the expressions for $\{\bE_0, \bE_b, \bK_b\}_{b=1}^{N}$ follow directly. For example,
\begin{align*}
\bE_b = \der{\bmeta_{\mu}}{\by_b} \bH_{\mu} \der{\bmeta_{\mu}}{\by_b^T} + \der{\bmeta_{\theta}}{\by_b} \bH_{\theta} \der{\bmeta_{\theta}}{\by_b^T},
\end{align*}
substituting the results of \eqref{first}-\eqref{thetab} into $\bE_b$ and after some algebra, it can be shown that
\begin{align*}
\bD_{E,b} & \approx \frac{c^2\Pn\lambda_b\beta_1}{2\beta_b}
    \begin{bmatrix}
    \delta_1^2 + \delta_2^2 & -(\delta_1+\delta_2) \\
    -(\delta_1+\delta_2) & 2
    \end{bmatrix}  + \frac{c^2\Pn\epsilon_b\beta_1}{2\beta_b^3}
    \begin{bmatrix}
    2f_b^2& 0 \\
    0 & 0
    \end{bmatrix} ,
\end{align*}
where $\delta_1 = \T_{1,b}-\Omega_1/\beta_1$, $\delta_2 = \T_{2,b}+\T_{1,2}\gamma_{2,b}-\Omega_2/\beta_2$, and we have approximated $\gamma_{1,b}=\gamma_{2,b}=\gamma_{1,2}\approx 1$.
The expression for $\bbA_{E,b} - \bbB_{E,b}\bbD_{E,b}^{-1}\bbB_{E,b}^T$ follows as
\begin{align}
\bbA_{E,b} - \bbB_{E,b}\bbD_{E,b}^{-1}\bbB_{E,b}^T & = \frac{c^2\Pn}{4} \mathbb{E}_{\bbeta}\begin{bmatrix}
                      \lambda_b\beta_b &    -\lambda_b\beta_b &        0 &               0\\
                     -\lambda_b\beta_b &     \lambda_b\beta_b &        0 &               0\\
0 &   0 & \epsilon_b/\beta_b  &                   -\epsilon_b/\beta_b \\
                   0 & 0 &  -\epsilon_b/\beta_b & \epsilon_b/\beta_b
\end{bmatrix}\nonumber\\
&- \frac{c^2\Pn\lambda_b}{4}\mathbb{E}_{\bbeta}\begin{bmatrix}
  \eta_1^2 &      \eta_1\eta_2&        \zeta_1/f_b &                         \delta_1/f_b\\
                      \eta_1\eta_2&      \eta_2^2&       -\zeta_2/f_b &               \delta_2/f_b\\
 \zeta_1/f_b &   -\zeta_2/f_b &  -\Omega_b\zeta_1/f_b^2&                   -\Omega_b\delta_1/(\beta_bf_b^2) \\
              \delta_1/f_b & \delta_2/f_b &  -\Omega_b\delta_1/(\beta_bf_b^2) & 0
\end{bmatrix},
\end{align}
where $\eta_1 = \Wb\delta_1/(\Tb f_b)$, $\eta_2 = \Wb\delta_2/(\Tb f_b)$, $\zeta_1 = \Omega_b(1-\eta_1^2)/\beta_b+\delta_1$, $\zeta_2 = \Omega_b(1+\eta_1\eta_2)/\beta_b-\delta_2$. When Assumption \ref{assumption} holds, we have the result in \eqref{lemma:Eb}. 
\end{IEEEproof}

\bibliographystyle{IEEEtran}
\bibliography{IEEEabrv,Asynchronous_Localization}

\end{document}